\documentclass[%
 preprint,
unsortedaddress,
 amsmath,amssymb,
 aps,
prc,
]{revtex4-1}

\usepackage{color}
\usepackage{graphicx}
\usepackage{dcolumn}    
\usepackage{multirow}
\usepackage{bm}         
\usepackage{sidecap}
\usepackage{hyperref}   
\usepackage{xcolor}
\usepackage{amsmath}
\usepackage{color,colortbl}
\usepackage{booktabs} 
\usepackage{hhline}
\usepackage{ulem}
\usepackage{hyperref} 
\usepackage{physics}
\usepackage{footnote}

\definecolor{dark-red}{rgb}{0.,0.,0}
\definecolor{dark-blue}{rgb}{0.,0.,1}
\definecolor{medium-blue}{rgb}{0,0,1}
\definecolor{gray}{rgb}{0.85,0.85,0.85}

\hypersetup{
    colorlinks, linkcolor={dark-red},
    citecolor={dark-blue}, urlcolor={medium-blue}
}

\begin{document}

\title{Statistical Hauser-Feshbach model description of $(n,\alpha)$ reaction cross sections for the weak s-process}%

\author{Sema K{\"u}\c{c}{\"u}ksucu}   
\affiliation {Department of Physics, Faculty of Science, University of Zagreb, Bijeni\v{c}ka c. 32, 10000 Zagreb, Croatia}%
\affiliation {Department of Physics, Faculty of Arts and Science, Aksaray University, 68100 Aksaray, Turkey}
\author{Mustafa Yi\u{g}it}
\affiliation{Department of Physics, Faculty of Arts and Science, Aksaray University, 68100 Aksaray, Turkey}
\author{Nils Paar}
\email{npaar@phy.hr}
\affiliation{Department of Physics, Faculty of Science, University of Zagreb, Bijeni\v{c}ka c. 32, 10000 Zagreb, Croatia}%

\date{\today}%

\begin{abstract}
The $(n,\alpha)$ reaction  contributes in many processes of energy generation and nucleosynthesis in stellar environment. Since experimental data are available for a
limited number of nuclei and in restricted energy ranges, at present only theoretical studies can provide predictions for all astrophysically relevant $(n,\alpha)$ reaction cross sections. The purpose of this work is to study $(n,\alpha)$ reaction cross sections for a set of nuclei contributing in the weak s-process nucleosynthesis. 
Theory framework is based on the statistical Hauser-Feshbach model implemented in TALYS code with nuclear masses and level densities based on Skyrme energy density functional.
In addition to the analysis of the properties of calculated $(n,\alpha)$ cross sections, the Maxwellian averaged cross sections are described and analyzed for the range of temperatures in stellar environment. Model calculations determined astrophysically relevant energy windows in which $(n,\alpha)$ reactions occur in stars. In order to reduce the 
uncertainties in modeling $(n,\alpha)$ reaction
cross sections for the s-process, novel experimental studies are called for. Presented results 
on the effective energy windows for $(n,\alpha)$ reaction in weak s-process 
provide a guidance for the priority energy ranges in the future experimental 
studies.
\end{abstract}

\maketitle

\section{Introduction}
Fascinating phenomena in the universe, such as evolution of stars, supernova explosions and neutron-star mergers, as well as the synthesis of chemical elements that occur in stellar environment, crucially depend on  a variety of nuclear reactions \cite{Burbidge1957, Cameron1957,Janka2007,Jose2011}. 
Nuclear reaction cross sections and rates are an essential for understanding the elemental abundances in our solar system and the galaxy. Since the properties of many nuclear reactions and involved nuclei in stellar environment still remain beyond the reach of the most
advanced experimental facilities, theoretical modeling is necessary to provide relevant cross sections for astrophysical applications \cite{Jose2011,NUPECC2017}.  Available nuclear reaction cross sections, 
both from the experiment and model calculations are summarized in various data bases \cite{Otuka2014,ENDFB7.1,Rauscher2000,JEFF-3.3,JENDL4.0,ROSFOND,CENDL3.1}.

The stellar burning phases of stars build chemical elements up to the Iron group by sequences of fusion reactions \cite{NUPECC2017}. However, elements heavier than Iron are produced by other types or nuclear processes, in different stellar environment \cite{Jose2011}. Of particular importance are neutron capture reactions that govern two key processes in nucleosynthesis: (i) slow neutron capture process, known as s-process, which takes place e.g., in giant stars \cite{Bao1987,Kappeler2011} and rotating massive metal poor stars \cite{Banerjee2019}, and (ii) rapid neutron capture process, known as r-process which occurs in explosive stellar environments such as supernova \cite{Mathews1990,Cowan1991,Wanajo2003,Thielemann2011} and neutron star mergers \cite{Freiburghaus1999,Goriely2011,Pian2017,Kajino2019}.
The description of neutron capture reactions crucially depends on the nuclear structure and excitation properties of target and daughter nuclei involved \cite{Litvinova2009}. One of
the key quantities is the reaction $Q$ value, which in general can have both positive and
negative values, and the reaction outcome is governed by the energy of the incoming 
neutrons, i.e. by the respective astrophysical conditions which determine the 
neutron energy distribution. Advanced theoretical modeling of nuclear properties and respective neutron induced reactions is essential to provide reliable input for astrophysical applications \cite{Jose2011}.

{The focus of this study is the} $(n,\alpha)$ reaction in nuclei of relevance for the s-process
nucleosynthesis \cite{Pignatari2010}. For example, $^{41}\text{Ca}(n,\alpha)^{38}\text{Ar}$ is
considered as the most important neutron-induced reaction on $^{41}$Ca in stellar
conditions of the weak s-process at temperature about 300 MK \cite{Vermote2012}, 
i.e., it strongly dominates over the neutron induced $\gamma$-emission \cite{Wooseley1978}.
The $(n,\alpha)$ reaction  
contributes in many processes of energy generation and nucleosynthesis. 
It is also an important reaction in primordial and stellar 
CNO cycle \cite{Schatz1993}.
In order to determine reaction rates in stellar 
environment, the reaction cross section as a function of neutron energy is 
an essential ingredient \cite{Bertulani2016}. Although the experimental studies provide some of
the necessary cross sections, for systematical implementation of $(n,\alpha)$ reactions in 
nucleosynthesis calculations theoretical modeling of the cross sections is indispensable. 
{As pointed out in Ref. \cite{JEFF3.1}, limited amount of experimental data for 
$(n,\alpha)$ reaction in the 1 keV–1 MeV region, restricts comprehensive data analysis.}
Only small changes in the neutron capture reaction cross sections could have 
significant implications on the path of the nuclear processes that govern the synthesis 
of chemical elements \cite{Dan2019}. Therefore, it is necessary to investigate 
$(n,\alpha)$ cross sections from various theoretical models, with different assumptions and
microscopic nuclear properties in order to provide complementary description that could
also have implications on the s-process network calculations \cite{Pignatari2010}.

{The experimental data on $(n,\alpha)$ reaction cross sections for the s-process
are available for a limited set of nuclei and in restricted ranges of neutron energies \cite{Konobeyev1996,Forrest1986,Otuka2014}.}
Recent measurements of $(n,\alpha)$ reactions include e.g.,Refs.\cite{Fessler1998,Smet2007,Gledenov2000,Goeminne2000,Vermote2012,Weiss2014,Fotiades2015,Barbagallo2016,Gledenov2018,Praena2018,Bai2019}.
The $(n,\alpha)$ reaction has been measured in the s-process branching point 
at $^{59}$Ni, in view of considerable differences between model predictions and 
experimental data \cite{Weiss2014,Helgesson2017}. Novel experimental techniques and
detector systems have been recently developed to provide accurate new data \cite{Weiss2013,Gyurky2019}.
The NICE-detector opened new perspectives to determine 
neutron capture cross-sections with charged particle in the exit channel with 
sufficient accuracy, for different nuclear and astrophysical applications \cite{NICE2020}.
%
Presently available experimental data and the corresponding empirical formulae cannot 
provide complete cross sections necessary for applications in astrophysical and nucleosynthesis models. Therefore, in this work we aim to investigate
$(n,\alpha)$ reaction cross sections from the theory side, based  
on Hauser-Feshbach statistical model \cite{Hauser1952,Moldauer1975}
implemented in the nuclear reaction program TALYS-1.95 \cite{Koning2008,Koning2015}.
When possible, nuclear properties needed for this study are based on the
energy density functional theory, using Skyrme-type functional with improved description of pairing, HFB-17 \cite{Goriely2009}. Due to its relevance in the s-process nucleosynthesis, the
$(n,\alpha)$ reactions will also be analyzed for the range of temperatures 
characteristic in stellar environment, by averaging the cross sections over the 
Maxwell-Boltzmann distribution.

The Gamow peak represents one of the most important concepts in the study of
thermonuclear reactions in stars \cite{Newton2007,Glorius2019,Ciani2021}. The corresponding energy range known 
as Gamow window, determines the effective stellar energy region in which most charged-particle
induced nuclear reactions occur \cite{Newton2007,Rauscher2010,Fallis2020,Glorius2019}. 
%
In Ref.\cite{Rauscher2010} the Gamow windows
of a variety of astrophysical reaction rates have been explored in a systematic study over the
nuclide map using Hauser-Feshbach model.
%
Since available experimental
data on $(n,\alpha)$ reaction cross sections for s-process nuclei are incomplete,
predictions of astrophysically relevant energy ranges are of importance for the future
design and planning of nuclear astrophysics measurements, including those using 
radioactive ion beams at novel facilities. Therefore, one of the objectives of this 
work is to determine the astrophysically relevant energy 
window where the $(n,\alpha)$ reactions occur in stellar environment. In this way, we aim to provide 
 the guidelines for the energy ranges
to be measured in future experiments on s-process nuclei, that are necessary to reduce
currently existing systematic uncertainties in model calculations.

The paper is organized as follows. 
Section \ref{Sec2} includes a brief overview of the theory framework and methods 
for $(n,\alpha)$ reactions and their implementation in TALYS code. 
In Sec. \ref{Sec31} the results on the cross sections for the s-process nuclei are presented
and discussed. The results for astrophysically relevant energy ranges for 
$(n,\alpha)$ reactions in s-process nuclei and respective Maxwellian averaged 
cross sections are also given and discussed in Sec. \ref{Sec32} and \ref{Sec33}. 
Summary and conclusions are given in Sec. \ref{Sec4}.

\section{Theory framework \label{Sec2}}

Neutron induced reaction cross sections in this work are described using the 
statistical Hauser-Feshbach model implemented in the nuclear reaction program TALYS-1.95 \cite{Koning2008,Koning2015}. More details about the formalism of Hauser-Feshbach
statistical model are given in Refs. \cite{Hauser1952,Moldauer1975,Rauscher2000}. 
The model assumes the validity of the
compound nucleus reaction mechanism and a statistical distribution of nuclear excited states \cite{Moldauer1975,Kappeler2011}. 
This method is appropriate when the
level density in the contributing energy window around the peak of the projectile energy distribution is sufficiently high 
to justify a statistical treatment \cite{Rauscher2000}. The compound nucleus formation dominates when the energy of the incident particle is low enough, below $\approx$ 20 MeV. This condition is almost always satisfied in astrophysical environments. Among various outgoing particles obtained from the compound nucleus, dominating contributions to the cross section come from neutron, $\gamma$-ray, proton, and $\alpha$-particle.
Comprehensive Hauser-Feshbach model calculations of astrophysical reaction rates are available in
Ref. \cite{Rauscher2000}.

In the present work, nuclear reaction code TALYS-1.95 is used to calculate $(n,\alpha)$ reaction cross sections \cite{Koning2008,Koning2015}. This computational framework includes necessary input on nuclear structure properties, optical models, level densities, fission properties, etc. \cite{Koning2015}. Nuclear level densities along with optical model transmission coefficients are considered as two most important ingredients of the
statistical model. The knowledge on all relevant physical quantities and properties of target nuclei, could be used from available experimental data or theoretical models. However, in the case of nuclei for which experimental data are limited, theoretical models represent the only possible source of relevant nuclear data. Since many statistical model ingredients are not available from the experiment, phenomenological models are often used instead \cite{Rauscher2000}. 

The nuclear properties needed for this study, in particular nuclear masses and level densities, are calculated using the nuclear energy 
density fuctional (EDF) theory.  This framework represents the most complete description of ground-state properties, collective excitations and processes over the whole nuclide chart, from relatively light systems to superheavy nuclei, and from the valley of stability to the drip-lines \cite{Stoitsov2009,Paar2005,Paar2010,Khan2011,Samana2011,Fantina2012,Moustakidis2014,Niksic2014,Paar2007,Gao2013,Xavi2013,Washiyama2012,Yuksel2019}.
Among the microscopic approaches to the nuclear many-body problem, no other method achieves comparable global accuracy at the same computational cost.  It is the only approach that can describe the evolution of nuclear structure throughout the nuclide chart. In practical implementations the nuclear EDF framework is analogous to Kohn-Sham density functional theory (DFT) \cite{Kohn1965,Kohn1999}. It is widely used method for electronic structure calculations in condensed matter physics and quantum chemistry. Energy density functionals have so far been constructed mostly empirically, with their parameters usually adjusted to properties of symmetric and asymmetric nuclear matter, and bulk properties of a set of nuclei. 

{For the purpose of the present work, two main approaches based on the TALYS code are employed \cite{Koning2008,Koning2015}. In the first case the experimental masses are used, and the level densities are based on a Fermi gas model. We denote this setting as TALYS-a. The second approach is more consistent, based on microscopic theoretical description, that employs the Skyrme-type functional for the description of nuclear masses and level densities
(in the following, denoted as TALYS-b). In particular, the Skyrme functional with improved description of pairing, HFB-17, is used. More details on the interaction parameters, constrained within Hartree-Fock-Bogoliubov model are available in Ref. \cite{Goriely2009}.} The Skyrme EDF has been in the past successfully employed in description of a variety of nuclear properties and astrophysically relevant processes. 
{As pointed out in Ref. \cite{Goriely2008}, when
dealing with nuclear astrophysics applications, various nuclear inputs should be, when
possible, determined from global, universal and microscopic models. Therefore, the implementation of the Skyrme functional represents a reasonable and consistent approach in modeling the $(n,\alpha)$ reaction cross sections.}
{The Hauser-Feshbach model also requires description of the transmission 
coefficients for the $\alpha$ particle emission. In this work we employ well established
$\alpha$ optical model 
potential from Ref. \cite{Avrigeanu2014}, where the parameterisation used are given
in detail in the TALYS-1.95 implementation \cite{Talys}. Compound nucleus cross section also includes width fluctuation correction (WFC) which accounts for the correlations between the 
incident and outgoing waves \cite{Talys}. These correlations enhance the elastic channel, and accordingly decrease other open channels. In this work the WFC factors are calculated
using the Moldauer model \cite{Moldauer1980}.}
With these settings, we use the TALYS-1.95 code \cite{Talys} to calculate the $(n,\alpha)$ reaction 
cross sections, as well as the Maxwellian averaged cross sections (MACS) of particular relevance for the weak s-process nucleosynthesis and future
experimental studies of $(n,\alpha)$ reactions at novel research facilities.

\section{Results and Discussion \label{Sec3}}

\subsection{$(n,\alpha)$ reaction cross sections for s-process nuclei \label{Sec31}}

\begin{figure}
\includegraphics[width=10.5 cm]{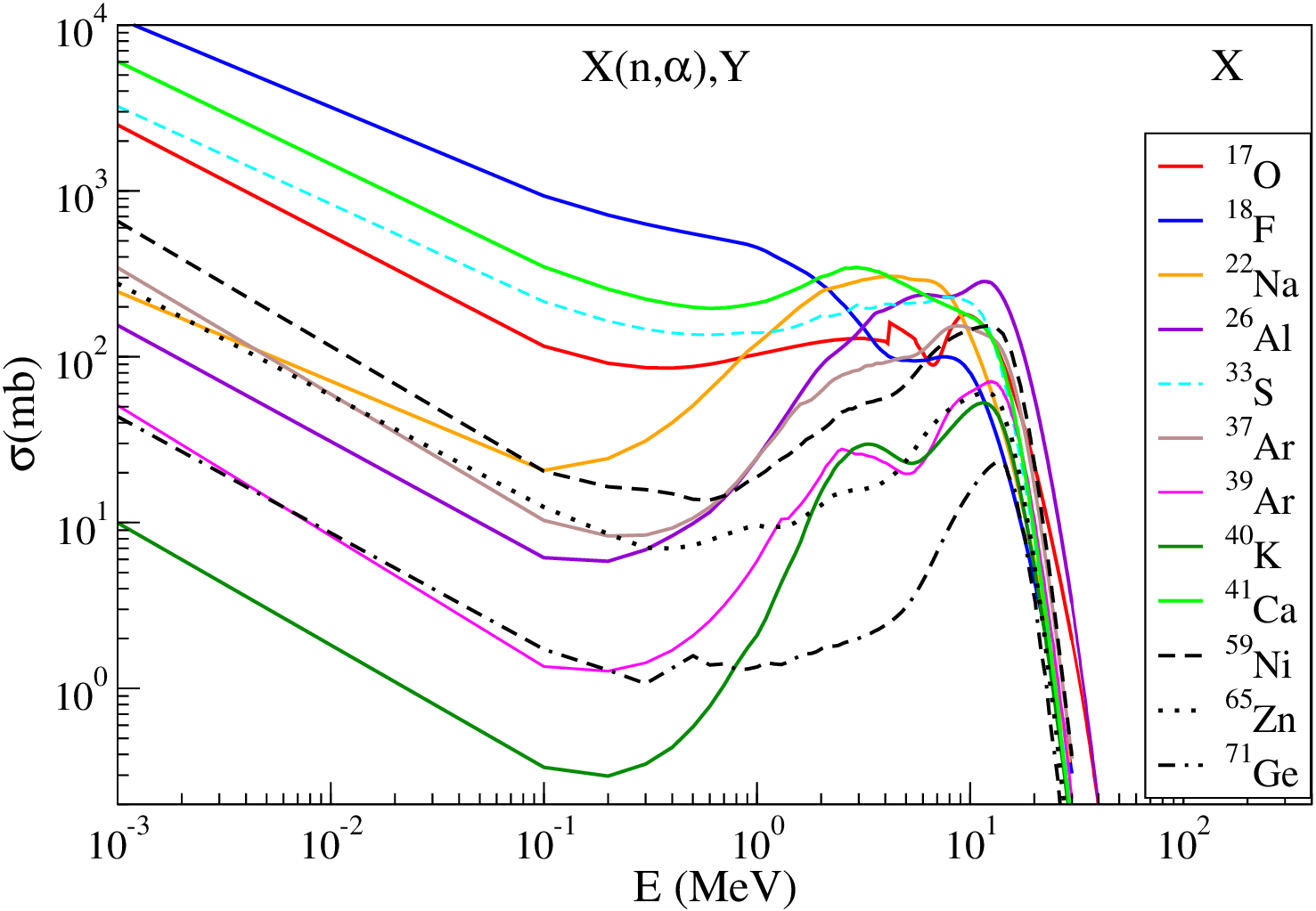}
\caption{The $(n,\alpha)$ reaction cross section as a function of the incoming neutron energies for target nuclei as listed in the figure. \label{fig-s-nuclei}}
\end{figure}

The theory framework outlined in Sec.\ref{Sec2} is employed {in the study of 
$(n,\alpha)$ reaction cross sections for nuclei introduced in the weak s-process network 
calculations of a massive star, including convective He burning core
and shell C- burning \cite{Pignatari2010}.  The weak s-process produces most of the
s-process isotopes between iron and strontium \cite{Pignatari2010}. As pointed out
in Ref. \cite{Pignatari2010}, the s-process nucleosynthesis is rather uncertain, 
especially in the C shell, due to uncertainties in the neutron capture cross 
sections. Therefore, in this study we aim to
improve the knowledge on the $(n,\alpha)$ reaction cross sections of importance
for the weak s-process, with focus on twelve relevant nuclei as introduced
in Ref. \cite{Pignatari2010}. These nuclei,
together with the corresponding $(n,\alpha)$ reaction Q-values (in MeV) are:
$^{17}$O (1.817) ,$^{18}$F (3.442), $^{22}$Na (1.9523), $^{26}$Al (2.462), $^{33}$S (3.870), $^{37}$Ar (3.264), $^{39}$Ar (3.060), $^{40}$K (2.973), $^{41}$Ca (6.101), $^{59}$Ni (5.711), $^{65}$Zn (6.223), and $^{71}$Ge (5.704).  We note that for all nuclei the Q values 
are calculated using Skyrme  HFB-17 interaction, except for $^{17}$O and $^{22}$Na,
where the experimental values are adopted since the model calculations resulted
in larger differences when compared to the experiment (we use these values also 
in the forthcoming calculations of the cross sections). For all nuclei under 
consideration, the Q values are positive,
thus allowing $(n,\alpha)$ reaction already at low neutron energies, that is of particular
relevance for the respective reaction rates in stellar environment, as will be discussed
in Sec. \ref{Sec32}. Figure \ref{fig-s-nuclei} shows the $(n,\alpha)$ reaction cross sections 
for the set of twelve nuclei introduced above.}
For the sake of completeness, we present not only astrophysically
relevant low-energy range of the cross sections, {emphasized by using logarithmic
scale, but a complete result. At low energies,  $E<$100 keV, the cross sections show the
$1/v$ dependence that is a general feature in neutron induced reactions in this energy range \cite{Krane1988}.} 
At the intermediate energies, all of the cross sections are peaked within the energy range 
$E\approx$ 1$-$20 MeV. Their maximal values vary from $\approx$30 mb for $^{71}$Ge 
toward more than 300 mb for $^{41}$Ca and $^{18}$F. The shape of the cross
sections corresponds to the general expectation for the compound nucleus reaction,
i.e. the $(n,\alpha)$ reaction cross section increases to a maximum and then 
decreases because higher energies start opening new emission channels \cite{Krane1988}.
The details of the cross sections sensitively depend
on the properties of nuclei involved in reactions, in particular on their masses, excitation spectra and
level densities. Clearly, the variations of the cross sections for presented nuclei indicate the 
necessity for systematic calculations that include all relevant nuclear properties. 
%
%

\begin{figure}
\includegraphics[width=12.5 cm]{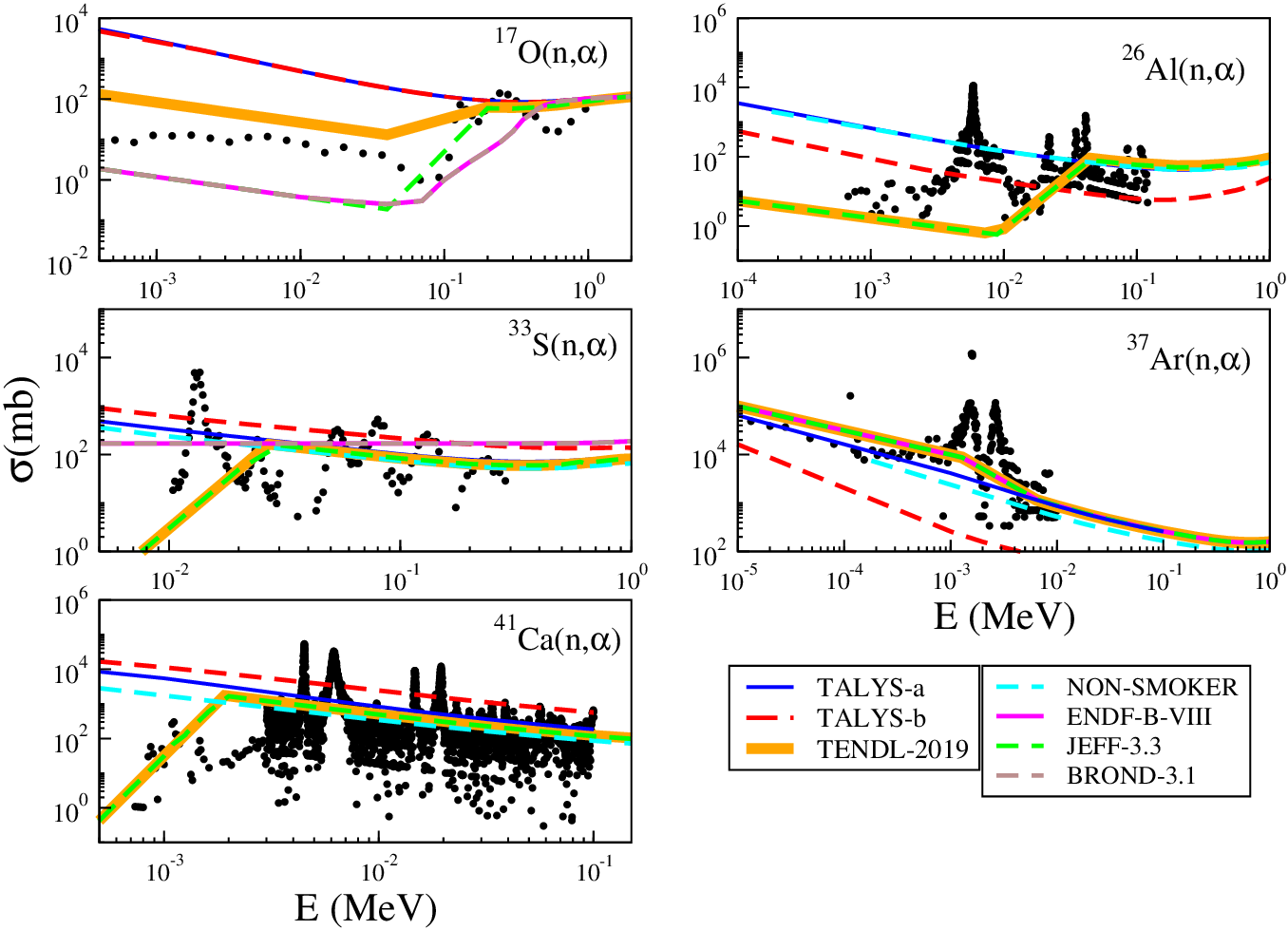}
\caption{Comparison between the calculated and experimental $(n,\alpha)$ reaction cross sections for $^{17}$O, $^{26}$Al, $^{33}$S, $^{37}$Ar, and $^{41}$Ca. Model calculations include 
experimental nuclear masses and level densities from the Fermi gas model (TALYS-a) and
nuclear masses and level densities from the Skyrme functional (TALYS-b). 
For comparison, results from the 
NON-SMOKER \cite{Rauscher2000} TENDL-2019 \cite{TENDL-2019}, ENDF-B-VIII \cite{ENDF-B-VIII}, JEFF-3.3 \cite{JEFF-3.3} and BROND-3.1 \cite{BROND3.1} data 
sets are also shown. The experimental data are taken from Refs. \cite{Smet2007,Koehler1995,Wagemans1987,Vermote2012,Goeminne2000}.
\label{Low-energies}}
\end{figure}

Model calculations are especially important for $^{18}$F, $^{22}$Na, $^{39}$Ar, $^{40}$K, $^{65}$Zn and $^{71}$Ge, 
because no experimental data are existing on $(n,\alpha)$ reactions for these nuclei, or they
are very limited.
For other s-process nuclei studied in this work, some experimental data are available
in the restricted low-energy range \cite{Koehler1991,Smet2007,Koehler1995,Wagemans1987,Goeminne2000,Vermote2012,Weiss2014}.
Figure \ref{Low-energies} shows the $(n,\alpha)$ reaction cross sections 
for {$^{17}$O, $^{26}$Al, $^{33}$S, $^{37}$Ar, and $^{41}$Ca} target nuclei, where the experimental data exist in the low-energy range\cite{Smet2007,Koehler1995,Wagemans1987,Vermote2012,Goeminne2000}. Two different 
Hauser-Feshbach model calculations have been performed, (i) with experimental masses and level densities from the Fermi gas model (TALYS-a) and (ii) nuclear masses and level densities calculated with the Skyrme functional (TALYS-b), with the exception of $^{17}$O, where the experimental mass is used, due to the reasons discussed above. In this way one can also obtain the information on the sensitivity of the cross sections on these two essential ingredients in modeling reaction cross sections, i.e. up to an order of magnitude difference is obtained between the results of
these two calculations. Nevertheless, the cross sections correspond reasonably well to
the experimental data for {$^{26}$Al \cite{Smet2007}, $^{33}$S \cite{Koehler1995,Wagemans1987}, $^{37}$Ar \cite{Goeminne2000}, and $^{41}$Ca \cite{Vermote2012}.} For comparison, Figure \ref{Low-energies} 
also shows the results from the evaluated data sets 
NON-SMOKER \cite{Rauscher2000} TENDL-2019 \cite{TENDL-2019}, ENDF-B-VIII \cite{ENDF-B-VIII}, JEFF-3.3 \cite{JEFF-3.3} and BROND-3.1 \cite{BROND3.1}.
These data sets include results of Hauser-Feshbach model calculations,
however, in comparison to the present work, they are based on different selections
of nuclear masses, excitation spectra, level densities, optical potential, WFC factors, 
and as a result the corresponding cross sections are subject to variations depending 
on the nuclear input.
Figure \ref{Low-energies} shows that in most of the cases, reasonable agreement 
is obtained between the present results and experimental data. Comparison with other
evaluated data display qualitative agreement though some systematic differences 
are obtained, except at lower-end energies where the TENDL-2019
data show rapid decrease of the cross sections for several orders of magnitude (except 
for $^{37}$Ar). There is no reported explanation for the strong kink observed for the
TENDL-2019 cross sections.
Although experimental data also exist for the $^{59}$Ni cross section \cite{Weiss2014,Helgesson2017}, these are not shown here because of a single
resonance peak structure obtained at 203 eV excitation energy, for which the
Hauser-Feshbach model is not applicable. For $^{40}$K, only a single data point
is available, at very low energy, 0.025 eV. Thus, novel experimental studies are needed
over a broader astrophysically relevant energy range that would also allow to constrain and improve model calculations.

\subsection{Astrophysically relevant neutron energy window for $(n,\alpha)$ reactions \label{Sec32}}

\begin{figure}
\includegraphics[width=10.5 cm]{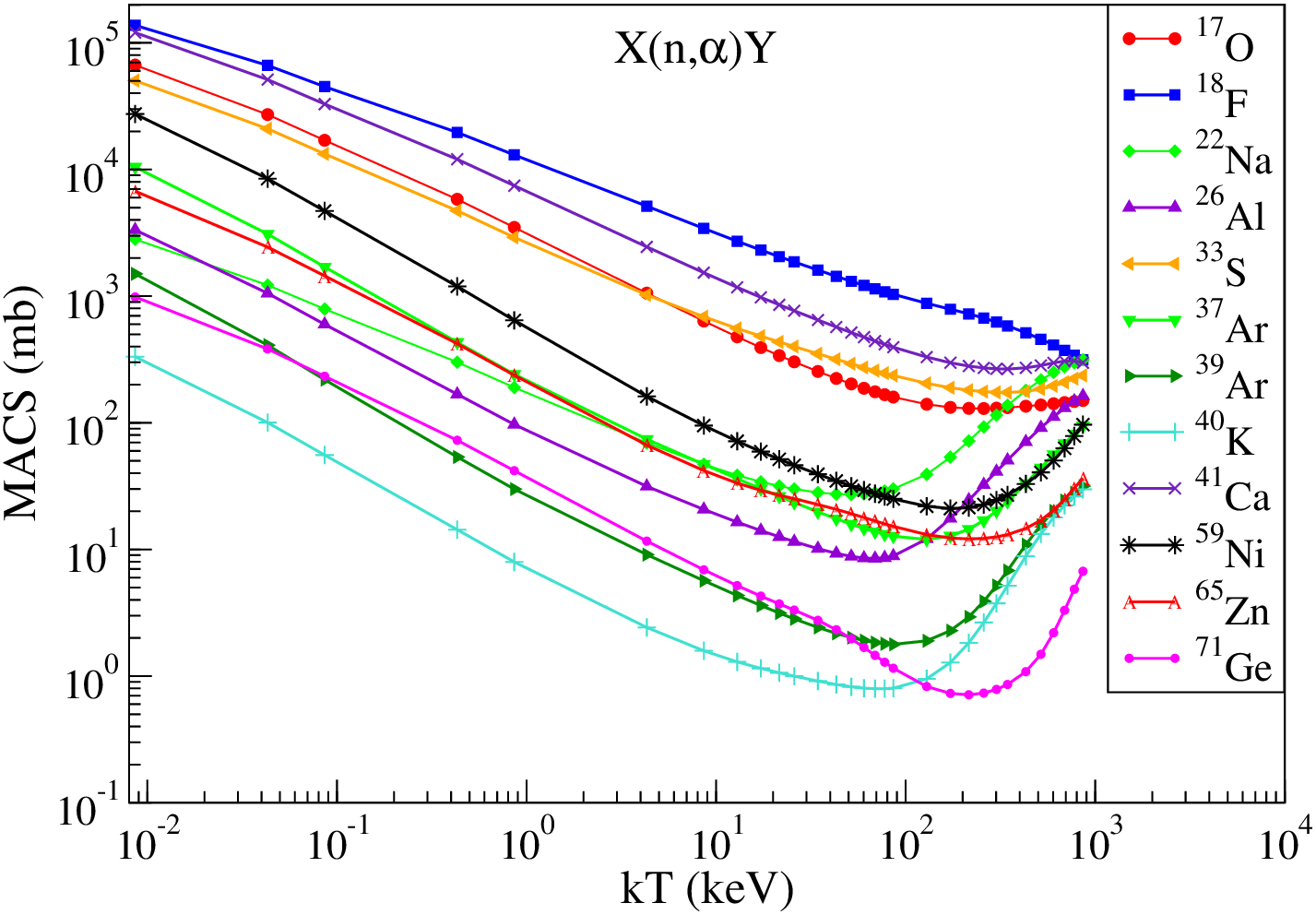}
\caption{The $(n,\alpha)$ reaction cross sections averaged over the Maxwell-Boltzmann distribution for the set of nuclei shown as a function of temperature.\label{maxwelliancs}}
\end{figure}

Next we explore the $(n,\alpha)$ reaction cross sections relevant for the s-process 
nucleosynthesis in stellar environment. For this purpose the cross sections are averaged 
over the Maxwell-Boltzmann distribution, that describes the distribution of the neutron energies with respect to the corresponding temperature \cite{Pritychenko2010},
\begin{equation}
\langle \sigma \rangle(kT)=\frac{2}{\sqrt{\pi}}(kT)^{-2} \int_{0}^{\infty} \sigma(E)Ee^{-\frac{E}{kT}}dE,
\label{macs}
\end{equation}
where $k$ and $T$ denote the Boltzmann constant and temperature, respectively.
Figure \ref{maxwelliancs} shows the Mawellian averaged $(n,\alpha)$ reaction 
cross sections  (MACS) \cite{Pritychenko2010} obtained for the same set of  
nuclei discussed in Sec. \ref{Sec31}. 
In the astrophysically relevant low-temperature range, one can observe a 
systematic decrease of the MACS values, that is governed by the $1/v$ dependence
of the cross sections shown in Figure \ref{fig-s-nuclei}. Following the Gaussian-like shape of
the cross sections at intermediate energies, $E\approx$1$-$30 MeV (see Figure \ref{fig-s-nuclei}),
the corresponding MACS values start to increase in the temperature range $kT\approx$100keV $-$ 1 MeV.
Model calculations also identify a hierarchy of the MACS values, showing the largest cross sections
for the four target nuclei, $^{18}$F, $^{41}$Ca and $^{33}$S, and $^{17}$O that reflect large cross sections in the low energy region shown in Figure \ref{fig-s-nuclei}. The MACS values for other nuclei appear
in most of the cases an order of magnitude smaller. The resulting behaviour of the MACS values
could have implications in network calculations of the weak s-process nucleosynthesis, identifying relevant
contributions from $(n,\alpha)$ reactions for nuclei where the corresponding cross sections have
larger values. However, these cross sections need to be considered in competition with
neutron induced reactions with other possible exit channels.%

\begin{figure}
\includegraphics[width=10.5 cm]{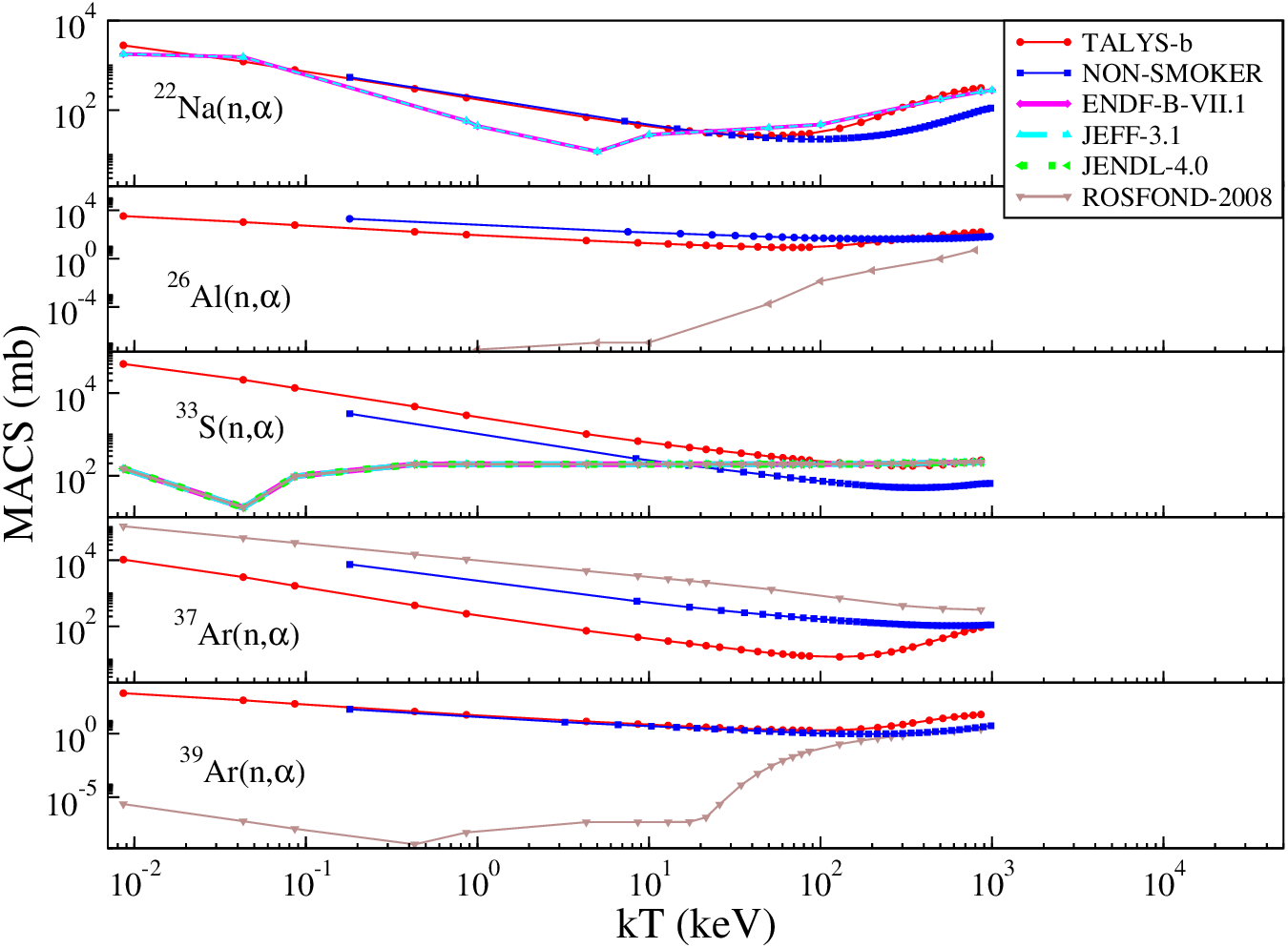}
\caption{The Maxwellian averaged cross sections (MACS) for $(n,\alpha)$ reaction for ${}^{22}$Na,  ${}^{26}$Al, and  ${}^{33}$S, ${}^{37}$Ar, and ${}^{39}$Ar  as functions of temperature. The present results (TALYS-b) are compared to those from NON-SMOKER \cite{Rauscher2000}, ENDF-B-VII.1 \cite{ENDFB7.1}, JEFF-3.1 \cite{JEFF3.1}, JENDL-4.0 \cite{JENDL4.0}, and ROSFOND-2008 \cite{ROSFOND}. \label{figNa-Ar}}
\end{figure}

In order to assess the information on the systematic uncertainties in modeling $(n,\alpha)$ reactions,
it is necessary to compare the results from various theoretical approaches based on different assumptions.
In Ref. \cite{Oginni2011} the variations of theoretical results have been explored for $^{26}\text{Al}(n,\alpha)^{23}\text{Na}$
reaction cross sections when using three different approaches, TALYS \cite{Koning2008}, EMPIRE \cite{Herman2007}, and NON-SMOKER \cite{Rauscher2000}. 
Figure \ref{figNa-Ar} shows the MACS results of the present work for $^{22}$Na,  ${}^{26}$Al, and  ${}^{33}$S, ${}^{37}$Ar, and ${}^{39}$Ar target nuclei in comparison
with those from several other model calculations based 
on different assumptions on the nuclear properties and approaches in modeling the reaction 
mechanism. These include NON-SMOKER reaction cross sections data base \cite{Rauscher2000}, 
Evaluated Nuclear Data File (ENDF-B-VII.1) \cite{ENDFB7.1}, Joint Evaluated Fission and Fusion (JEFF-3.1) 
Nuclear Data Library \cite{JEFF3.1}, JENDL-4.0 Library for Nuclear Science and Engineering \cite{JENDL4.0}, and
ROSFOND-2008 nuclear data \cite{ROSFOND}.
The MACS results from this work are in qualitative agreement 
with the Hauser-Feshbach calculation NON-SMOKER for $^{22}$Na (except for kT>100 keV),
${}^{26}$Al, and ${}^{39}$Ar, while for ${}^{33}$S and ${}^{37}$Ar even more than an order of magnitude difference is obtained. Agreement with the ENDF, JEFF and JENDL evaluation data 
depends on the range of $kT$ values of interest. For example, the MACS values for
 ${}^{33}$S in this work, as well as those from NON-SMOKER, are considerably larger
 than ENDF, JEFF, and JENDL evaluation data for $kT<$ 10 keV.
In the case of ${}^{26}$Al, ${}^{33}$S, and ${}^{39}$Ar, for $kT<$ 100 keV, large discrepancies are obtained between the present 
MACS values (as well as those from other models) and the ROSFOND evaluation data, which also show different trend with increasing temperature. Clearly, different foundations
in various model calculations result in considerable model dependence of the MACS values.

\begin{figure}
\includegraphics[width=10.5 cm]{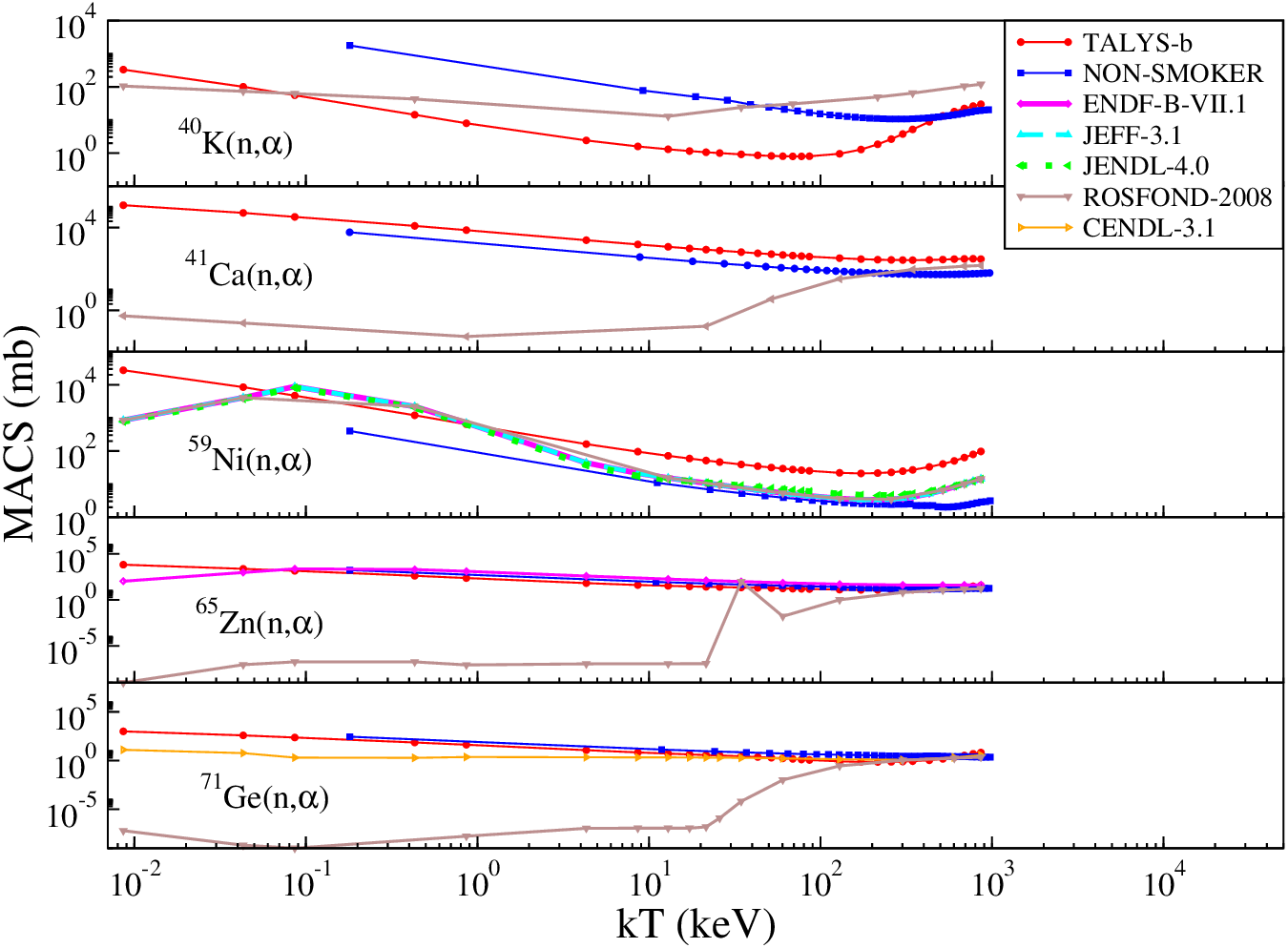}
\caption{The same as in Figure \ref{figNa-Ar}, but for 
$^{40}$K,  ${}^{41}$Ca, ${}^{59}$Ni, ${}^{65}$Zn, and ${}^{71}$Ge target nuclei. In addition to the results from other data sets, comparison with the MACS values from the CENDL-3.1 \cite{CENDL3.1} nuclear data library is also shown. \label{figK-Ge}}
\end{figure}

Figure \ref{figK-Ge} shows the MACS values for
$^{40}$K,  ${}^{41}$Ca, ${}^{59}$Ni, ${}^{65}$Zn, and ${}^{71}$Ge target nuclei.
In addition to the comparison with other evaluations as shown before, for $^{65}$Zn 
the results from the Chinese Evaluated Nuclear Data Library (CENDL-3.1) are also displayed. For all nuclei except $^{40}$K, good qualitative agreement of TALYS-b MACS values with other evaluations is obtained, except for the ROSFOND-2008 which
result in considerably smaller values except for $^{40}$K.
The differences obtained from comparison of the cross sections from various theoretical 
frameworks demonstrate the necessity for calculations of the MACS values using
different approaches and parameterizations, and analyses of their impact
in the network calculations of the s-process nucleosynthesis in realistic stellar conditions.

\subsection{Astrophysically relevant neutron energy range for $(n,\alpha)$ reactions \label{Sec33}}

\begin{figure}
\includegraphics[width=10.5 cm]{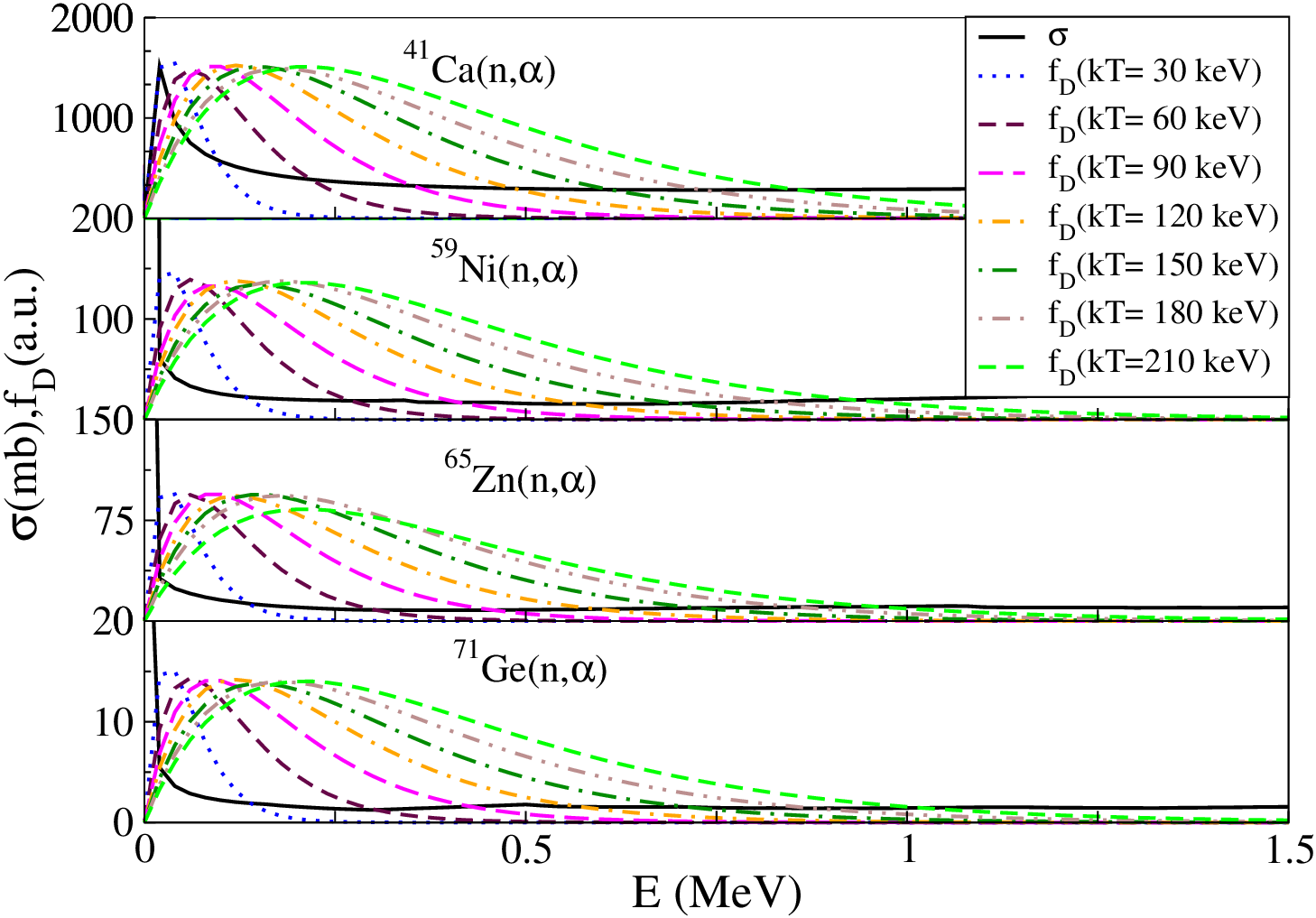}
\caption{The $(n,\alpha)$ reaction cross sections as a function of neutron energy $\sigma(E)$, and Maxwell-Boltzmann distribution
$f_D(E)$ (in arbitrary units) for the range of temperatures $kT=$ 30$-$210 keV, shown for 
${}^{41}$Ca, ${}^{59}$Ni, ${}^{65}$Zn, and ${}^{71}$Ge. \label{maxwell}}
\end{figure}

\begin{figure}
\includegraphics[width=10.5 cm]{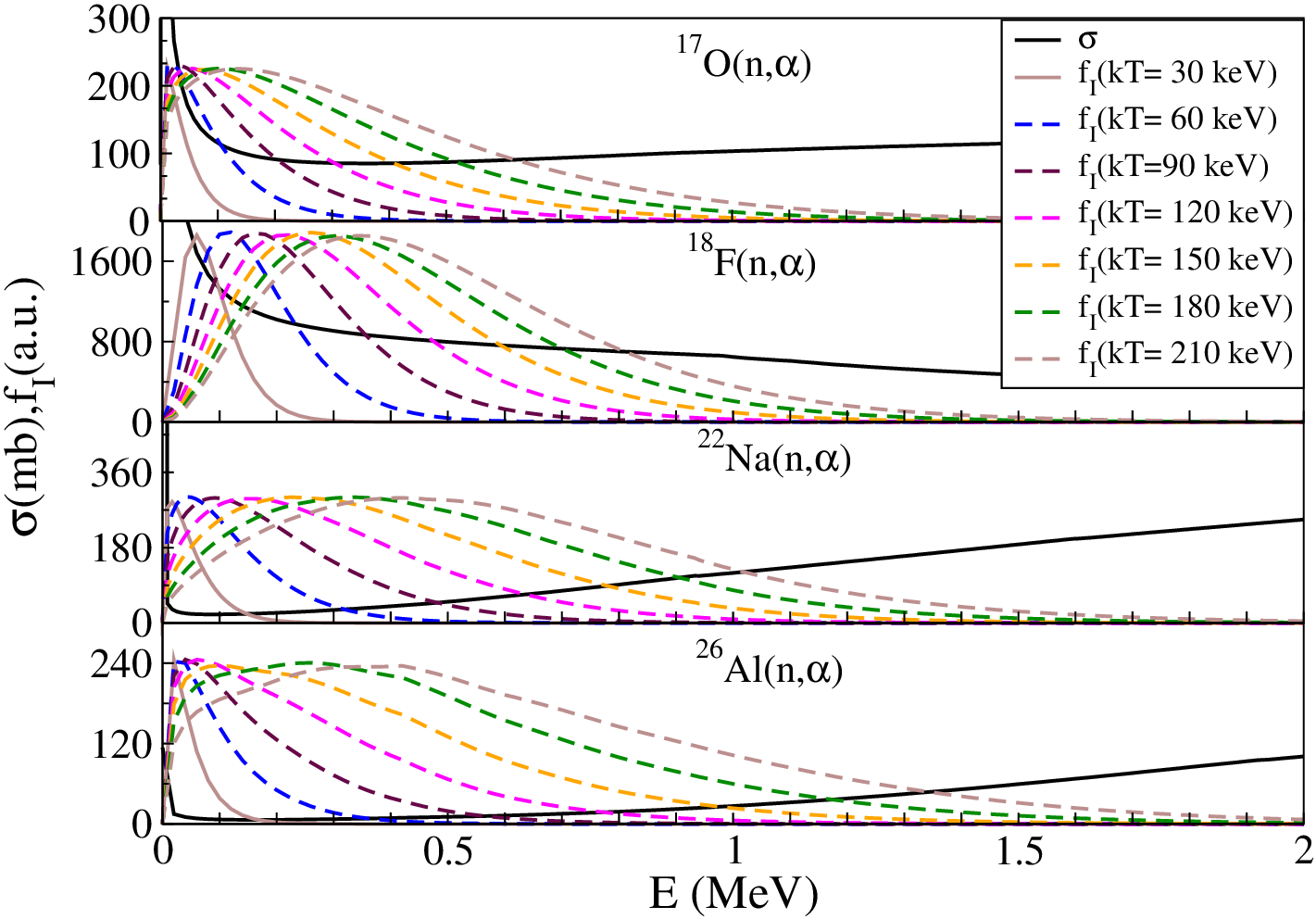}
\caption{Relevant $(n,\alpha)$ reaction energy windows for the range of temperatures
$kT=$ 30$-$210 keV described with the function $f_I(E)$ (in arbitrary units, see text) and the respective
reaction cross sections for ${}^{17}$O,${}^{18}$F, ${}^{22}$Na, and ${}^{26}$Al target nuclei. \label{1sema}}
\end{figure}

\begin{figure}
\includegraphics[width=10.5 cm]{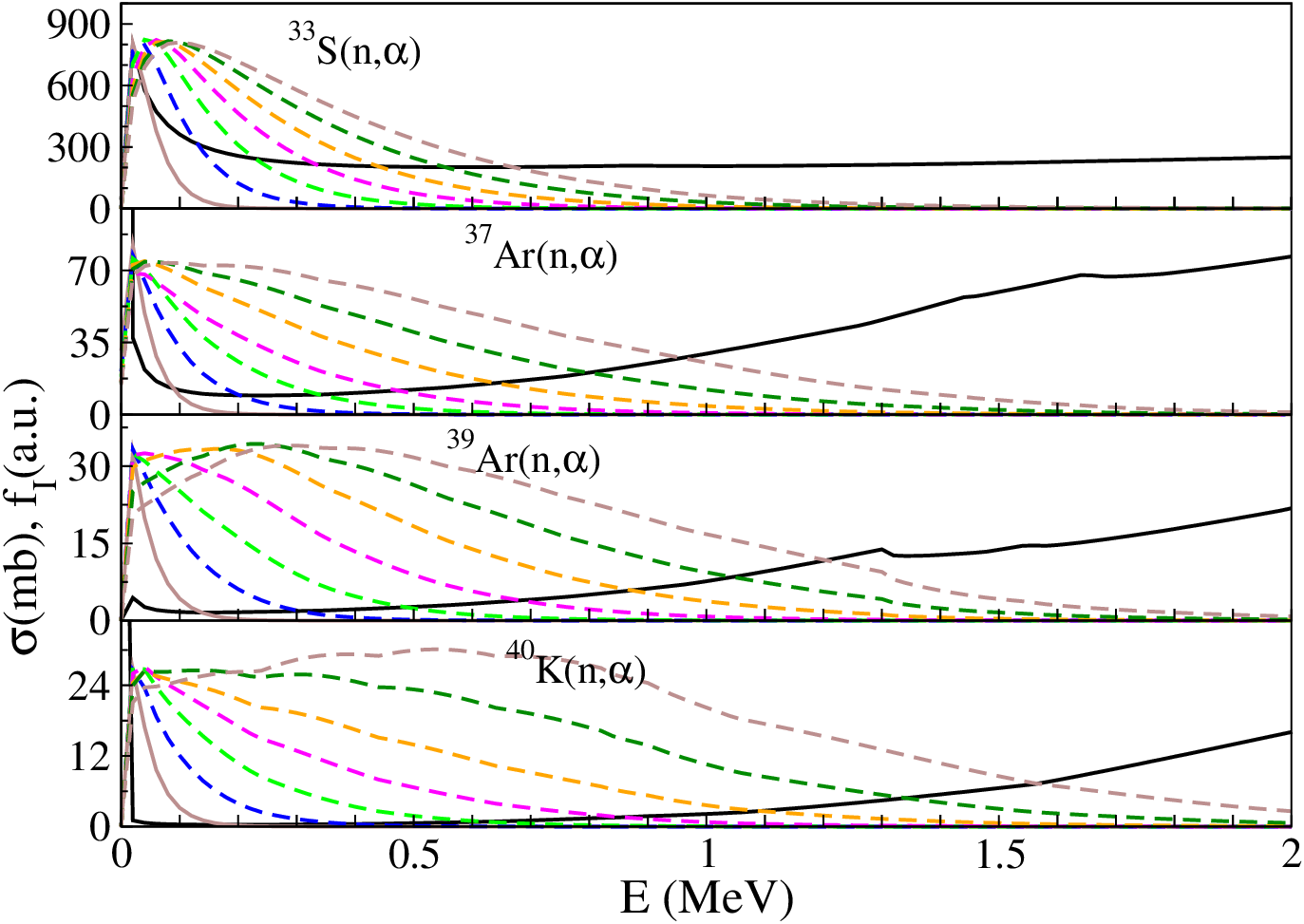}
\caption{The same as Figure \ref{1sema}, but for ${}^{33}$S, ${}^{37}$Ar, ${}^{39}$Ar, and ${}^{40}$K target nuclei. \label{2sema}}
\end{figure}

\begin{figure}
\includegraphics[width=10.5 cm]{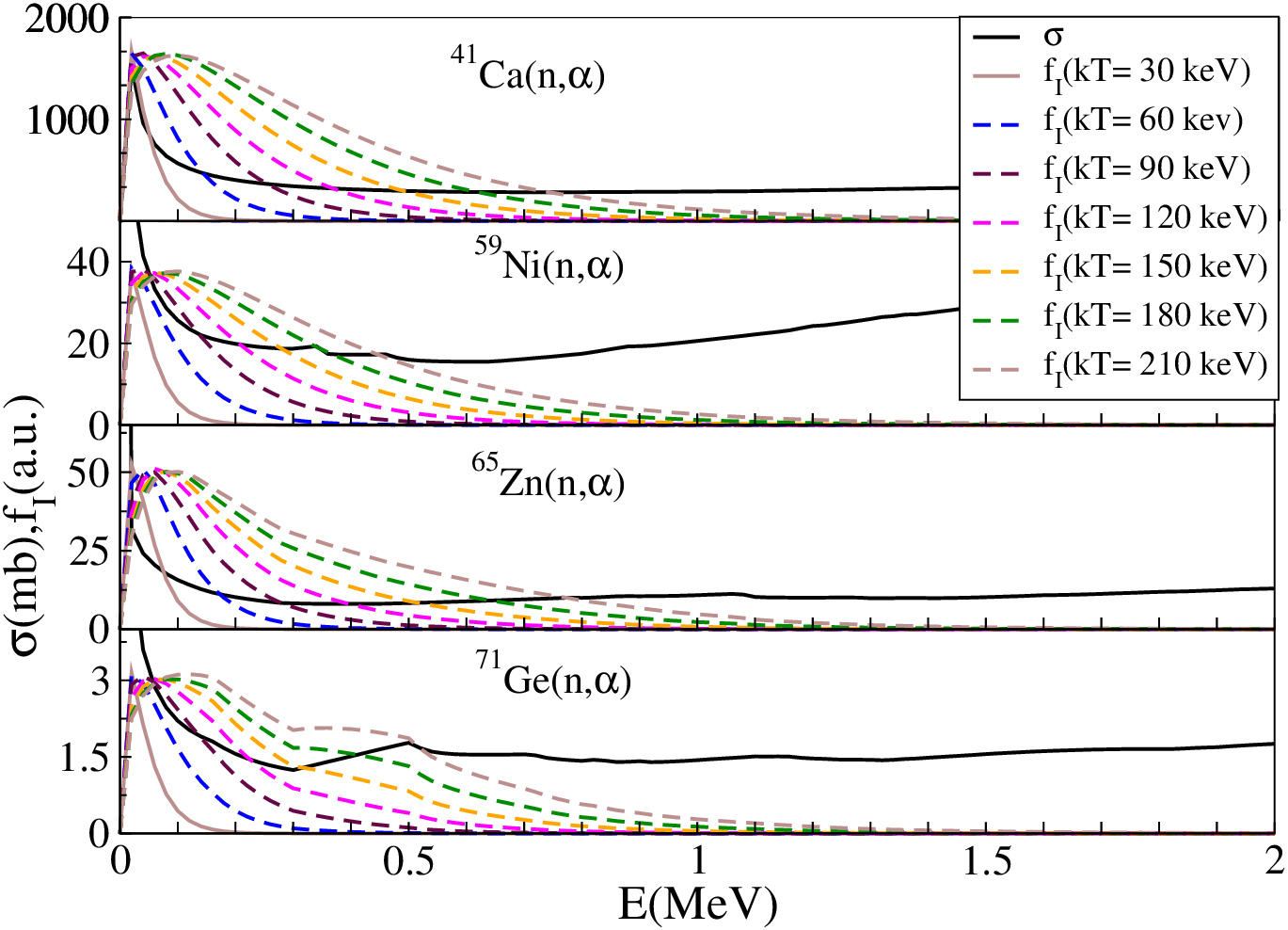}
\caption{The same as Figure \ref{1sema}, but for ${}^{41}$Ca, ${}^{59}$Ni, ${}^{65}$Zn, and ${}^{71}$Ge target nuclei. \label{3sema}}
\end{figure}

As shown in Sec. \ref{Sec32}, the $(n,\alpha)$ reaction cross sections are subject to a considerable
model dependence. In order to have more stringent constraints on the astrophysically relevant
neutron induced reactions, novel experimental studies are needed. As discussed in
Sec. \ref{Sec31}  available experimental data on s-process nuclei studied in this work 
are rather limited.
One of the objectives of this study is to determine what are the astrophysically
relevant energy ranges that could contribute in modeling the nucleosynthesis.
These relevant energy ranges, known as Gamow window, are determined as the 
overlap region between the Maxwell Boltzmann distribution of the interacting 
particles and the low-energy tail of the reaction cross section \cite{Newton2007}. The exact position
of the Gamow window depends on typical temperature regimes characteristic for
a specific core and burning stages during stellar evolution. 
While in the case of reactions that involve the charge particles, the Gamow peak is determined
by the interplay of the Maxwell-Boltzmann distribution and tunneling of the incoming particle
throught the Coulomb barrier of target nucleus \cite{Rauscher2010}, in the case of $(n,\alpha)$ reactions there 
is no Coulomb barrier for the incoming particle and the relevant energy range is determined
by the overlap of the Maxwell-Boltzmann distribution and the reaction cross section. This 
is illustrated in Figure \ref{maxwell} which shows the contributions from the Maxwell-Boltzmann
distribution $f_D(E)$ (in arbitrary units for the presentation purposes) and $(n,\alpha)$ reaction cross sections for  ${}^{41}$Ca,${}^{59}$Ni, ${}^{65}$Zn, and ${}^{71}$Ge as a function of neutron energy. 
The distributions are displayed for the range of temperatures $kT=$ 30$-$210 keV.
For these temperatures the Maxwell-Boltzmann distributions limit the neutron energy 
range at low energies. For comparison, the $(n,\alpha)$ 
cross sections are also shown. Thus, only the low-energy part of the cross sections are relevant 
for astrophysical applications within the temperature range as given above.
In the following we determine the relevant energy windows for $(n,\alpha)$ reactions for the complete previously studied set of nuclei. For this purpose, we analyze the subintegral function
used in calculating the MACS values (\ref{macs}),
\begin{equation}
f_I(E)= \sigma(E)Ee^{-E/(kT)},
\end{equation}
that provides the cross section weighted by the corresponding Maxwell-Boltzmann 
distribution at a given energy.
The reaction energy windows described by the function $f_I(E)$, together with the corresponding 
$(n,\alpha)$ reaction cross sections are shown in Figure \ref{1sema} for ${}^{17}$O,${}^{18}$F, ${}^{22}$Na,
and ${}^{26}$Al. As expected, the reaction energy windows become wider with increasing 
temperature, but their exact location sensitively depends on the specific target nucleus under consideration. For all nuclei, the maximal reaction peaks for $kT=\text{30 keV}$ are located below energy of 0.1 MeV.
To assess the accurate information on astrophysically relevant energy windows, 
calculations need to
be performed for each nucleus of interest. Therefore, we 
show the results for relevant $(n,\alpha)$ reaction windows also for ${}^{33}$S, ${}^{37}$Ar, ${}^{39}$Ar, and ${}^{40}$K in Figure \ref{2sema} and for ${}^{41}$Ca, ${}^{59}$Ni, ${}^{65}$Zn, ${}^{71}$Ge
in Figure \ref{3sema}. The results give quantitative predictions which energy windows are important
for different nuclei within given temperature range, and in this way provide
a guidance for the future experimental studies of $(n,\alpha)$ reactions, and their implementation in the s-process network calculations. As already discussed in 
Sec. \ref{Sec32}, considerable model dependence of  $(n,\alpha)$  reaction cross sections necessitate more experimental data that could provide additional constraints on these reactions 
that contribute to the s-process nucleosynthesis.

\section{Conclusions \label{Sec4}}

In this work we have investigated $(n,\alpha)$ reaction cross sections for the set of nuclei 
contributing to the s-process nucleosynthesis. Model calculations have been performed 
in the theory framework based on Hauser-Feshbach statistical model through its
implementation in the nuclear reaction program TALYS-1.95 \cite{Koning2008}.
When possible, model calculations are kept consistent by using nuclear properties
necessary for the reaction calculated in the framework of Skyrme energy density
functional. The $(n,\alpha)$ reactions have been investigated in view of their relevance
for the weak s-process nucleosynthesis, thus their average over the Maxwell-Boltzmann
distribution for the range of stellar temperatures has been calculated.

An important contribution of this work is a quantitative assessment of the energy window
in which $(n,\alpha)$ reactions occur in stars, that is by its role equivalent to the well known
Gamow window, with a difference that in the case of $(n,\alpha)$ reactions no tunneling
through the Coulomb barrier is needed in the entrance channel. Model calculations
in this work determined relevant energy windows for $(n,\alpha)$ reactions for the
range of astrophysically represented temperatures, indicating a strong dependence of 
the exact location of the energy window on specific target nucleus under consideration.
Quantitative predictions of astrophysically relevant energy windows for $(n,\alpha)$ reactions,
that contribute in the s-process nucleosynthesis, provide a guidance 
for the future experimental studies of these reactions. In particular, measurements 
in the energy windows as identified in Figures \ref{1sema}, \ref{2sema} and \ref{3sema} are necessary, especially for nuclei for which none or very limited data exist, $^{18}$F, $^{22}$Na, $^{39}$Ar, $^{40}$K, $^{59}$Ni, $^{65}$Zn, and $^{71}$Ge.
 As shown in this study,
$(n,\alpha)$  reactions, similar as all other neutron induced reactions, are subject to
a considerable systematic model dependence. Therefore, novel experimental data
are necessary in the predicted relevant energy windows, that could provide  
additional constraints on these reactions and reduce currently existing 
theoretical uncertainties. In the forthcoming studies the $(n,\alpha)$ reaction cross
sections obtained in this work could also be implemented in the s-process
network calculations to explore the variations of isotopes in the element abundances.

\section{Acknowledgements}
This work is supported by the QuantiXLie Centre of Excellence, a project co financed by the Croatian Government and European Union through the European Regional Development Fund, the Competitiveness and Cohesion Operational Programme (KK.01.1.1.01.0004). 
S.K. acknowledges support from the Scientific and Technological Research Council of Turkey (TUBITAK) through the International Doctoral Research Fellowship Programme 2214A, 2020/1, Grant No. 1059B142000254.

\bibliographystyle{apsrev4-1}
\bibliography{sprocess.bbl}

\end{document}